\documentclass[conference]{IEEEtran}
\IEEEoverridecommandlockouts
\usepackage{cite}
\usepackage{amsmath,amssymb,amsfonts}
\usepackage{algorithmic}
\usepackage{graphicx}
\usepackage{textcomp}
\usepackage{braket}
\usepackage{xcolor}
\usepackage{booktabs}
\usepackage{hyperref}
\usepackage{graphicx}
\usepackage{subfigure}

\begin{document}

\title{Implementing a Hybrid Quantum-Classical Neural Network by Utilizing a Variational Quantum Circuit for Detection of Dementia\\
}

\author{\IEEEauthorblockN{Ryan Kim}
\IEEEauthorblockA{\textit{Thomas Jefferson High School for Science and Technology} \\
Chantilly, USA \\
2024rkim@tjhsst.edu
}
}

\maketitle

\begin{abstract}
Magnetic resonance imaging (MRI) is a common technique to scan brains for strokes, tumors, and other abnormalities that cause forms of dementia. However, correctly diagnosing forms of dementia from MRIs is difficult, as nearly 1 in 3 patients with Alzheimer’s were misdiagnosed in 2019, an issue neural networks can rectify. The performance of these neural networks have been shown to be improved by applying quantum algorithms. This proposed novel neural network architecture uses a fully-connected (FC) layer, which reduces the number of features to obtain an expectation value by implementing a variational quantum circuit (VQC). 
This study found that the proposed hybrid quantum-classical convolutional neural network (QCCNN) provided 97.5\% and 95.1\% testing and validation accuracies, respectively, which was considerably higher than the classical neural network (CNN) testing and validation accuracies of 91.5\% and 89.2\%. Additionally, using a testing set of 100 normal and 100 dementia MRI images, the QCCNN detected normal and demented images correctly 95\% and 98\% of the time, compared to the CNN accuracies of 89\% and 91\%. With hospitals like Massachusetts General Hospital beginning to adopt machine learning applications for biomedical image detection, this proposed architecture would approve accuracies and potentially save more lives. Furthermore, the proposed architecture is generally flexible, and can be used for transfer-learning tasks, saving time and resources.

\end{abstract}


\section{Introduction}
With the field of quantum physics growing rapidly, more and more applications of quantum computing are entering the computer science field \cite{soares}. Quantum computing offers several benefits over classical computing, which include faster computations, a broader variety of problems to solve, and better performance for machine learning tasks \cite{golestan} \cite{yan}. The properties of encoding information through superposition and quantum entanglement are not present in classical computing and could provide new ways to solve specific tasks if harnessed correctly. However, current available quantum computers have limited qubits, which represent the superposition between two quantum states, and are prone to noise, which is intrinsic quantum error. The current state of quantum computing is often labeled the \emph{Noisy Intermediate Scale Quantum} (NISQ) era \cite{preskill}. 

Due to the advantages of quantum computing such as using significantly less features than classical neural networks, technologies such as the variational quantum circuit have been developed \cite{du}. Variational quantum circuits are a relatively new technology, invented by Peruzzo et al. 2014 \cite{peruzzo}. More fundamentally the variational quantum eigensolver, it has many applications in quantum machine learning, quantum chemistry, optimization, and other fields \cite{tilly} \cite{martin}. In chemistry, it solves for the ground state energy of the Hamiltonian, which gives lots of critical information about a system \cite{cerezo}. Many have applied variational quantum circuits to machine learning, as it has been shown to offer several benefits compared to its classical counterparts \cite{maheshwari} \cite{jager} \cite{chen} \cite{khairy} \cite{zhu} \cite{farhi}.

This novel hybrid quantum-classical neural network was tested utilizing an MRI dataset of brains with and without dementia, which holds relevance to this day. Dementia is a growing issue in the world, as over 55 million people in the world suffer from it \cite{dementia}. Dementia is the deterioration of cognitive function and can be caused by a variety of factors, such as head trauma, alcohol, Parkison's, Alzheimer's, and more. Magnetic resonance imaging (MRI) is one of the most effective ways of diagnosing dementia, but identifying whether a patient has a certain condition is difficult because the images are complex and require subjective deduction \cite{farisco}\cite{bhuva}. One of the new ways to detect a certain condition through MRI images is to use neural networks, which can be much more accurate and objective than humans \cite{sappagh}. My research attempts to enhance existing machine learning methods by applying a variational quantum circuit layer to a pre-existing network, in hopes of improving the accuracy of the network. While others have started to take advantage of quantum applications in machine learning, an original, angle-encoded ansatz that can be trained has not been applied to the binary classification of dementia in MRIs.

\section{Materials and Methods}

\subsection{Description of Dataset}
\begin{figure*}
    \centerline{\includegraphics[scale=1.2]{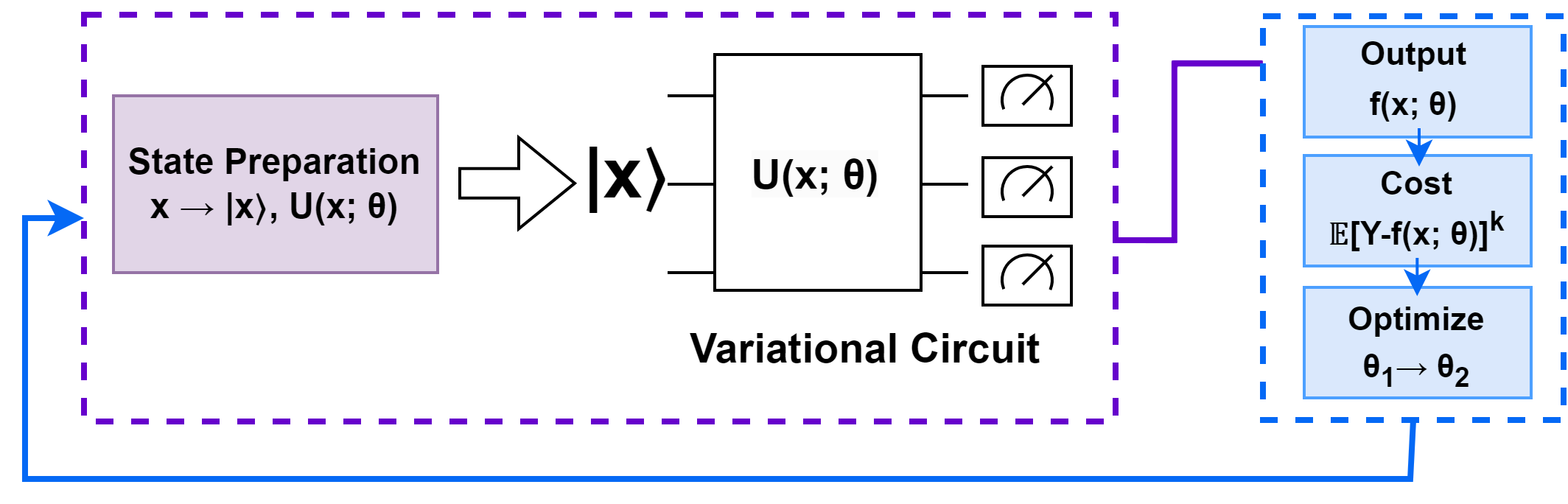}}
    \caption{Generalized variational quantum circuit, cost function, and optimization}
    \label{fig0}
\end{figure*}
The image dataset used in this study was produced and published publicly by Sarvesh Dubey on Kaggle \cite{dubey}. The dataset consists of both demented and non-demented MRI images of brains separated into four different classes representing different severities of dementia: not demented, mildly demented, moderately demented, and very demented. Overall, the total number of images was approximately 6400. 

\begin{figure}[htbp]
    
  \centering
  \begin{tabular}{ c @{\hspace{20pt}} c }
    \includegraphics[width=.3\columnwidth]{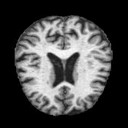} &
      \includegraphics[width=.3\columnwidth]{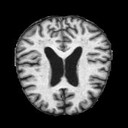} \\
    \small (a) &
      \small (b)
  \end{tabular}
    
  \medskip

  \caption{(a) Normal axial head of caudate and thalamus brain scan (b) Demented axial head of caudate and thalamus brain scan}
  \label{dementia}
\end{figure}

 Examples of images of the dataset can be seen in Figure \ref{dementia}, where a normal and demented brain can be seen on the left and right, respectively. One of the major differences between normal and demented brains is that the lateral ventricle is expanded abnormally, distorting the surrounding parts of the brain \cite{carimichael}.     

Since this study focuses on the binary classification between demented and non-demented images, I chose to use the 3199 non-demented and 2239 very demented images for my dataset \cite{dubey}.

\subsection{Preprocessing of Dataset}
The two pools of images were separated into different folders, and labeled 1 through the total number of images in the folder. By keeping the labels consistent within the folder, it becomes easy to iterate over the dataset to use for the model.

Additionally, images were resized to 250 x 250 pixels and the dataset array was converted into a Pytorch tensor.

\subsection{Practicality and Bias of Dataset}
While the image dataset used in the study is sufficient in size, much information is left out about the information of the human participants. Potential biases could have arrived with a lack of racial or gender diversity, both of which contribute to subtle differences within the brain matter \cite{studholme} \cite{liu}. Such biases cause the classification of the machine learning model to be biased, which is a major issue in machine learning \cite{alberto}. 

However, differences within the white brain matter, such as atrophies and other abnormalities can detected with the given dataset. Additionally, the main goal of study is to compare the advantage that hybrid quantum-classical neural networks have other their classical counterparts.

\subsection{Classical Neural Network}
A classical neural network consists of many different layers, used to extract features with mathematical functions. In deep-learning, feed-forward neural networks tend to be one of the most widely used models for object detection. A generalized layer of one feed-forward neural network can be represented by \cite{mari}:

\begin{equation}
    L_i: x_i \xrightarrow{} y_i = \psi (Wx_i + b)
\end{equation}

where $L_i$ is the specified layer, $W$ is the weights, $x_i$ is the input feature vector, $y_i$ is the output feature vector, b is the bias, and $\psi$ is a nonlinear activation function. The biases and weights are the values that will be optimized during the training process. 

The selected model used for feature extraction was the ResNet18, which is a sequential convolutional neural network, that includes 18 layers in total. ResNet18 consists of convolution, batch normalization, ReLU, and max pooling layers.

\subsection{Variational Quantum Circuit}

\subsubsection{Properties and Gates}

A variational quantum circuit takes advantage of certain quantum properties such as superposition, entanglement, and quantum gates to encode and manipulate data. A generalized version of the variational quantum circuit can be seen in Figure \ref{fig0}. Superposition is fundamental in quantum mechanics, as it is a system in multiple states, with each state having a corresponding probability, until a measurement occurs. Superposition of an arbitrary state in the z-basis can be represented by:
\begin{equation}
\ket{\psi} = a\ket{0} + b\ket{1}
\end{equation}

where $\ket{\psi}$ is the given state, \{$\ket{0}$, $\ket{1}$\} are the orthogonal Z-basis vectors, a is the corresponding probability amplitude of $\ket{0}$, and b is the corresponding probability ampltude of $\ket{1}$.

Entanglement is another foundational principle of quantum mechanics, which states that once measuring a state of one particle of an entangled pair, the state of the other particle is known, no matter how far the particles are from each other. A simple representation of an entangled system in a Stern-Gerlach experiment is:

\begin{equation}
\ket{\psi} = \frac{1}{\sqrt{2}}(\ket{0}_1 \ket{0}_2 + \ket{1}_1 \ket{1}_2)
\end{equation}

where \{$\ket{0}_1$, $\ket{1}_1$\} represent basis state vectors of the first particle and \{$\ket{0}_2$ ,$\ket{1}_2$\} represent basis state vectors of the second particle.

In the Z-basis, the spin-up and down vectors, $\ket{0}$ and $\ket{1}$, can be represented in matrix form as

\begin{equation}
    \ket{0} = \begin{pmatrix}
    1 \\
    0
    \end{pmatrix}
\end{equation}

\begin{equation}
    \ket{1} = \begin{pmatrix}
    0 \\
    1
    \end{pmatrix}
\end{equation}

\begin{figure}[htbp]
    \centerline{\includegraphics[scale=0.50]{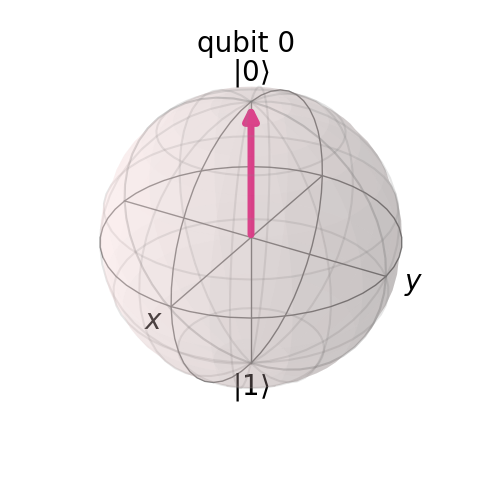}}
    \caption{Bloch sphere represented of a single qubit in the $\ket{0}$ state}
    \label{normal}
\end{figure}

In quantum computing, there are gates that perform operations on qubits, which store information. A single qubit in the $\ket{0}$ state can be represented visually as seen in Figure \ref{normal}.

\begin{figure}[htbp]
    \centerline{\includegraphics[scale=0.50]{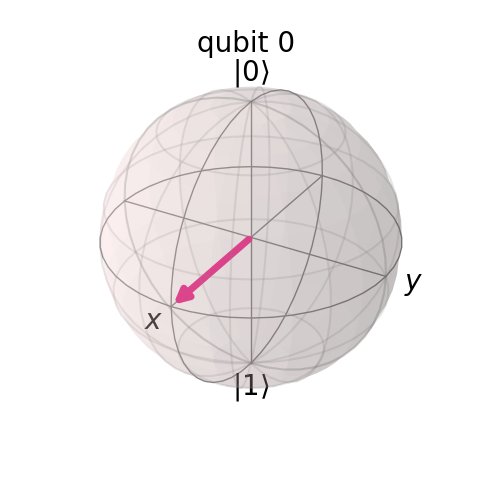}}
    \caption{Bloch sphere represented after applying the Hadamard gate to the $\ket{0}$ state}
    \label{hadamard}
\end{figure}

The most fundamental gate is the Hadamard (H) gate, which brings a single qubit into a superposition state. It can be represented in the matrix form as \cite{montanaro}
\begin{equation}
    H = \frac{1}{\sqrt{2}}\begin{pmatrix}
    1 & 1 \\
    1 & -1
    \end{pmatrix}
\end{equation}
A visual representation of the Hadamard after applying the gate to Figure \ref{normal} can be seen in Figure \ref{hadamard}.

\begin{figure*}
    \centerline{\includegraphics[scale=0.60]{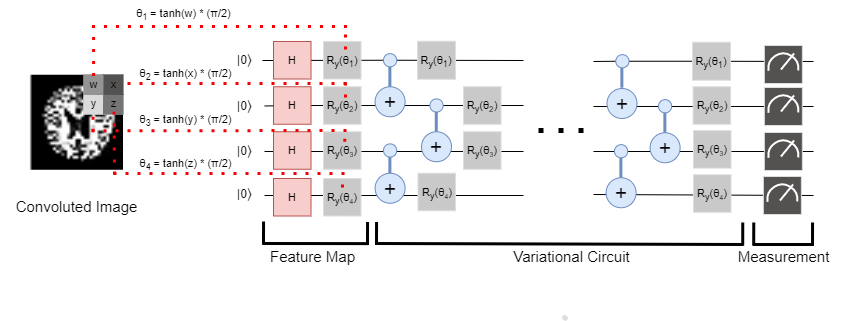}}
    \caption{Variational quantum circuit with embedding layer, variational layer, and measurement layer reducing 512 features to 4 features}
    \label{fig1}
\end{figure*}
Another important gate is the controlled-not (CNOT) gate, which checks if the control qubit has the state $\ket{1}$ and then flips the target qubit from $\ket{0}$ to $\ket{1}$ and vice versa. Its matrix representation is:

\begin{equation}
    CNOT = \begin{pmatrix}
    1 & 0 & 0 & 0 \\
    0 & 1 & 0 & 0 \\
    0 & 0 & 0 & 1 \\
    0 & 0 & 1 & 0
    \end{pmatrix}
\end{equation}

\begin{figure}[htbp]
    \centerline{\includegraphics[scale=0.50]{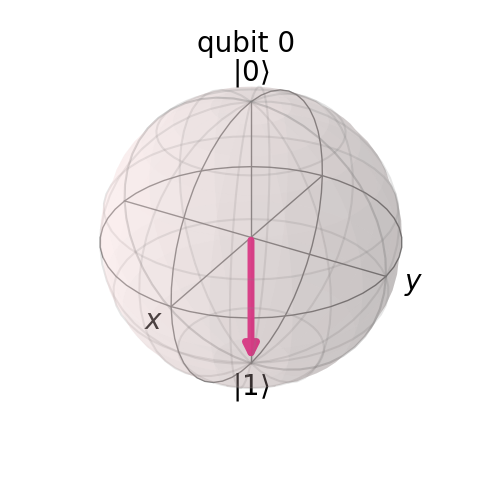}}
    \caption{Bloch sphere represented after applying the Rotation-Y gate to the $\ket{0}$ state}
    \label{ygate}
\end{figure}

The final gate used in this study is the rotation-y (RY) gate, which rotates a given qubit state a given number of radians around the complex y-axis. It can be represented as

\begin{equation}
    RY = \begin{pmatrix}
    cos(\frac{\theta}{2}) & -sin(\frac{\theta}{2})  \\
    sin(\frac{\theta}{2}) & cos(\frac{\theta}{2}) \\

    \end{pmatrix}
\end{equation}

where $\theta$ is the angle of rotation desired and Figure \ref{ygate} is the Bloch sphere representation.

In general, variational quantum circuits consist of three major layers: embedding, variational, and measurement layer. 

\subsubsection{Encoding}
The embedding layer will map a vector from the classical space to the quantum Hilbert space \cite{schuld}, and is parameterized by $\theta$, which is defined as

\begin{equation}
    \theta_{i, j}^{(k)} = tanh(I_{i, j}) * \frac{\pi}{2}
\end{equation}

where $I_{i, j}$ is the intensity of a pixel specified at ($i$, $j$) of a given convoluted input image, and k is the specified qubit number. In the four-qubit variational circuit used in this study, $k \in {1, 2, 3, 4}$ and the parameters at a given pixel specified at ($i$, $j$) are $\theta_{i, j}^{(k)}$.

The embedding layer $E$ consists of a Hadamard gate, and a Rotation-Y gate, which are applied to a qubit in the $\ket{0}$ state.

\begin{equation}
    E^{(k)} = R_{y}(\theta_{i, j}^{(k)})H
\end{equation}

\begin{equation}
    E: x_{k} \xrightarrow{} \ket{x}_{k} = E^{(k)}\ket{0}
\end{equation}

A variational layer, $L$, can be defined as \cite{mari}:
\begin{equation}
    L: \ket{\psi(x)} \xrightarrow{} \ket{\psi(y)} = U(w)\ket{\psi(x)}
\end{equation}
\begin{equation}
    \ket{\psi(x)} = \bigotimes_{k = 1}^{4} \ket{x}_k
\end{equation}
where $\ket{\psi(x)}$ is the four-qubit input state, $\ket{\psi(y)}$ is the output state, $w$ is a vector that consists of all the theta parameters, and U is a unitary gate matrix that represents the matrix product of the H, CNOT, and RY gates.

In machine learning, a common technique for classification is to non-linearly map to a higher dimension to solve for the hyperplane, which separates the data \cite{maheshwari}. In terms of the variational quantum circuit, it can be represented as \cite{maheshwari} \cite{mari}

\begin{equation}
    \label{unitary}
    U(w) = T \bigotimes_{k = 1}^{4}R_{y}(w_{k})\ket{\psi(x)}
\end{equation}

where $k$ is the specified qubit, $w_{k}$ is the specified variational parameter of vector $w$, and $\ket{x}$ is the state from the last layer.

T in equation \ref{unitary} represents the unitary entanglement operation, represented by

\begin{equation}
    T = C_{1,2} C_{3,4} C_{2,3}
\end{equation}

where the first $C$ subscript represents the control qubit and the second subscript represents the target qubit for CNOT gates.

The concatenation of these variational layers of depth d can be expressed as:
\begin{equation}
    Q = U_d \circ \cdots U_2 \circ U_1
\end{equation}

The final layer produces four classical outputs by measuring the quantum states $\ket{\psi_{k}(y)}$ from the four qubits and computes the expectation value $y$. The measurement operator $M$ can be expressed as:
\begin{equation}
    M: \ket{\psi(x)} \xrightarrow{} y = |\braket{\psi(x)|\psi(y)}|^2
\end{equation}

Overall, the variational quantum circuit V can be expressed as:
\begin{equation}
    V = M \circ Q \circ E
\end{equation}
where E is the embedding layer, Q is the variational circuit, and M is the measurement layer in the Z-basis.
\begin{table*}
\caption{}
\centering
\begin{center}
\begin{tabular}{cccccc}
\multicolumn{6}{c}{Hyperparameters for the hybrid and classical models} \\ \hline
Epochs & Learning Rate & Loss Function & Optimizer & Batch Size & Activation Function \\
\textbf{20} & \textbf{$10^{-4}$} & \textbf{Cross-Entropy} & \textbf{Adam} & \textbf{32} & \textbf{RELu} \\ \hline
\end{tabular}
\end{center}
\label{tab1}
\end{table*}
\subsection{Creating a Hybrid Quantum-Classical Neural Network}
\begin{figure}[htbp]
    \centerline{\includegraphics[scale=0.60]{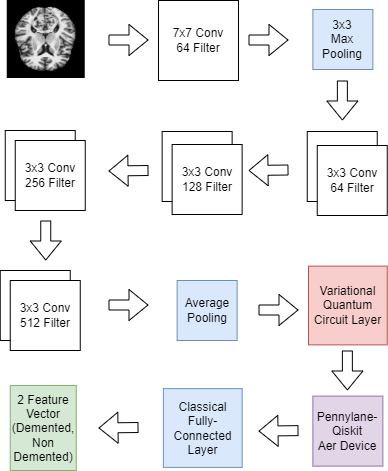}}
    \caption{Architecture of the Hybrid Quantum-Classical Neural Network}
    \label{fig2}
\end{figure}

The ResNet-18 neural network model was used the extract features from the input images, and the variational quantum circuit significantly reduces the number of features, which catalyzes the network towards classification. According to Figure \ref{fig2}, 512 features are extracted from the initial ResNet-18 average pooling layer and then fed into the variational quantum circuit, which is depicted in Figure \ref{fig1}.

After the variational quantum circuit computes the given quantum logic gates and applies a z-measurement on each qubit, the four dimensional vector is sent to the classical fully-connected (FC) layer, to which it is converted into two values: an expectation value for a brain with dementia and an expectation value for a brain without dementia.

Representing the quantum-classical neural network mathematically as $QCNN$:

\begin{equation}
QCNN = L_{4\xrightarrow{}2} \cdot Q_{512\xrightarrow{}4} \cdot L_{512}
\end{equation}

where $L_{512}$ represents the ResNet-18 Layers that extract 512 features, $Q_{512\xrightarrow{}4}$ is the variational quantum circuit that reduces the 512 features to 4, and ${L_{4\xrightarrow{}2}}$ is the fully-connected layer.

\subsection{Defining Hyperparameters}
In machine learning, hyperparameters define the learning process before training the model. Hyperparameters cannot be changed during the training process of the neural network model \cite{yu} \cite{yang}.

The first hyperparameter to be defined as the number of epochs. Epochs are defined as the number of times the whole dataset is passed through the model. In this study, the number of epochs is set to 20 to ensure that the accuracies and losses are able to converge.

The next hyperparameter to be defined as the learning rate. In machine learning, the learning rate defines how much the model changes its weights with respect to the loss gradient function. Generally, the higher the learning rate, the faster the loss function converges and the lower the learning rate, the more intricately and accurately the loss function converges \cite{dong}.

Another hyperparameter to be defined is the batch size, which is the number of images to be sent through the model in one iteration. The more images used in a batch, the slower the model is trained \cite{nakamura}. However, like the learning rate, time is exchanged for potentially higher accuracies and lower losses. In this study, the chosen batch size was 32 images.

The layers of a neural network are made up of neurons, which can be activated, from which information can be sent to the next layer. The activation function, which determines whether or not a neuron activates, is another important hyperparameter. In this study, the chosen nonlinear activation function is the rectified linear unit (ReLU), which can be defined as \cite{mari}:
\begin{equation}
ReLU(x) = max(0, x)
\end{equation}

The loss function is another important hyperparameter that quantifies how far the model prediction is from the actual value. The chosen loss function for this study is the cross-entropy loss function for binary classification, which can be expressed as \cite{mari}:

\begin{equation}
    CL = -\sum_{n=1}^{2}y_nlog(P_n)
\end{equation}
where $y_n$ is the output value computed by the model and $P$ is the overall probability of the $n$th class.

The last hyperparameter is the optimizer, which is an algorithm to minimize the loss function. The selected optimizer for this study is the Adam optimizer \cite{kingma}, which can compute large amounts of data effectively.

All of the hyperparameters are represented in Table \ref{tab1}.

\subsection{Performance Metrics}
To compare the ResNet-18 and the proposed model, the accuracy, recall, precision, and F1-score metrics are used \cite{recall}:

\begin{equation}
    accuracy = \frac{TP + TN}{TP + TN + FP + FN}
\end{equation}
\begin{equation}
    recall = \frac{TP}{TP + FN}
\end{equation}
\begin{equation}
    precision = \frac{TP}{TP + FP}
\end{equation}
\begin{equation}
    {F_1} = \frac{2 * recall * precision}{recall + precision}
\end{equation}

where $TP$ represents the true positives, which are data values correctly given a positive label. $TN$ represents the true negatives, which are data values correctly given a negative label. $FP$ represents the false positives, which are data values incorrectly given a positive label. $FN$ represents the false negatives, which are data values incorrectly given a negative label.

For calculating the testing metrics, the same 100 unused non-demented and 100 unused demented brain MRIs were used. Conditions were kept the same to prevent confounding variables to influence the comparison of performance metrics between the two models.
\section{Results And Discussion}

\subsection{Training}
\begin{figure}[htbp]
    \centerline{\includegraphics[scale=0.40]{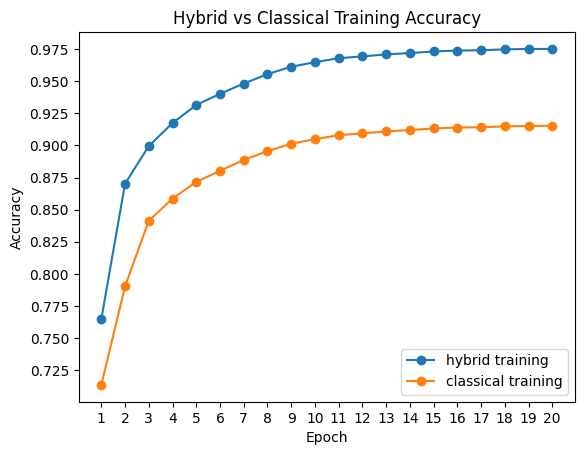}}
    \centerline{\includegraphics[scale=0.40]{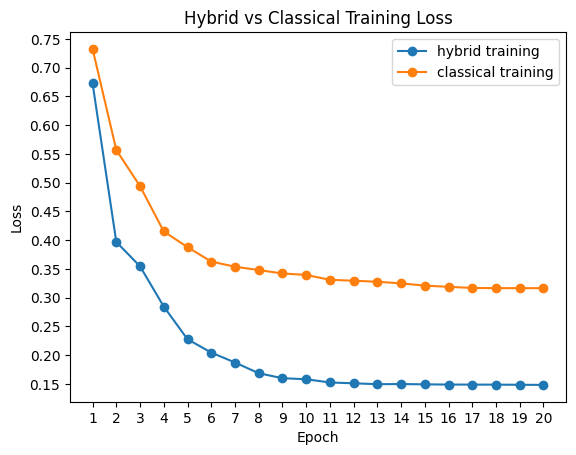}}
    \caption{Training accuracy and loss with respect to epoch number for both hybrid and classical models}
    \label{fig3}
\end{figure}

During the training phase, the accuracy and loss graphs for the classical and hybrid neural networks are shown in Figure \ref{fig3}.

In the top graph of Figure \ref{fig3}, the hybrid model consistently has a higher accuracy than the classical model, with the difference of accuracy during convergence being approximately 0.065. Both models drastically improve in accuracy between epochs 1-3. In the bottom graph of Figure \ref{fig3}, the hybrid model consistently has less loss than the classical model, with the difference of loss during convergence being approximately 0.2. Both models initially start with large values of loss and then drastically decrease in loss between epochs 1-5.

After both models reached convergence, the training accuracy was approximately 0.976 for the hybrid model and 0.915 for the classical model. The corresponding convergence training losses were 0.148 and 0.316 for the hybrid and classical models, respectively.

\begin{table*}
\caption{}
\begin{center}
\begin{tabular}{ccccccccc}
\multicolumn{9}{c}{Comparison  of classical to hybrid model} \\ \hline
Model & Training Accuracy & Validation Accuracy & Training Loss & Validation Loss & Testing Accuracy & Recall & Precision & F1-Score \\
\textbf{Hybrid} & \textbf{0.975} & \textbf{0.951} & \textbf{0.148} & \textbf{0.380} & \textbf{0.965} & \textbf{0.980} & \textbf{0.951} & \textbf{0.966} \\
Classical & 0.915 & 0.892 & 0.316 & 0.557 & 0.900 & 0.91 & 0.892 & 0.901 \\ \hline
\end{tabular}
\end{center}
\label{tab2}
\end{table*}
\subsection{Validation}
\begin{figure}[htbp]
    \centerline{\includegraphics[scale=0.40]{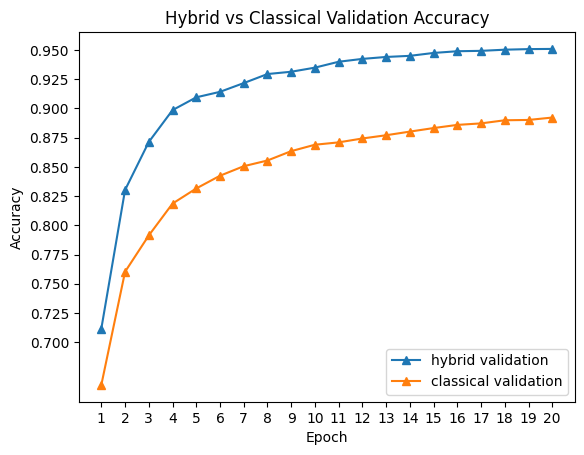}}
    \centerline{\includegraphics[scale=0.40]{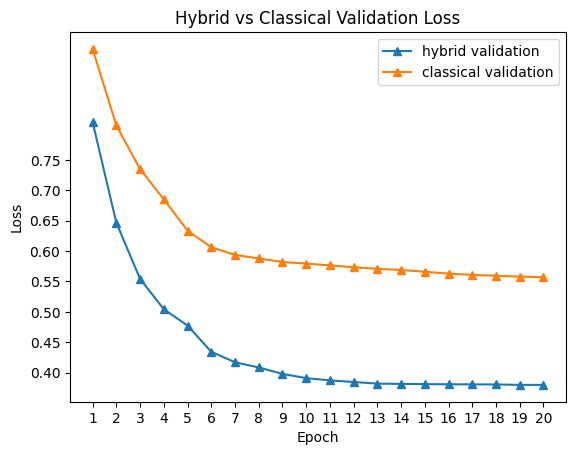}}
    \caption{Validation accuracy with respect to epoch number for both hybrid and classical models}
    \label{fig5}
\end{figure}

In the top graph of Figure \ref{fig5}, the hybrid model once again consistently has a higher accuracy than the classical model, with the difference of accuracy during convergence being approximately 0.06. Both models drastically improve in accuracy between epochs 1-4. In the bottom graph of Figure \ref{fig5}, the hybrid model consistently has less loss than the classical model, with the difference of loss during convergence being approximately 0.18. Both models initially start with large values of loss and then drastically decrease in loss between epochs 1-5, but less drastically than the training losses.

\subsection{Testing}
After both models reached convergence, the training accuracy was approximately 0.976 for the hybrid model and 0.915 for the classical model. The corresponding convergence training losses were 0.148 and 0.316 for the hybrid and classical models, respectively.

\begin{figure}[htbp]
    \centerline{\includegraphics[scale=0.32]{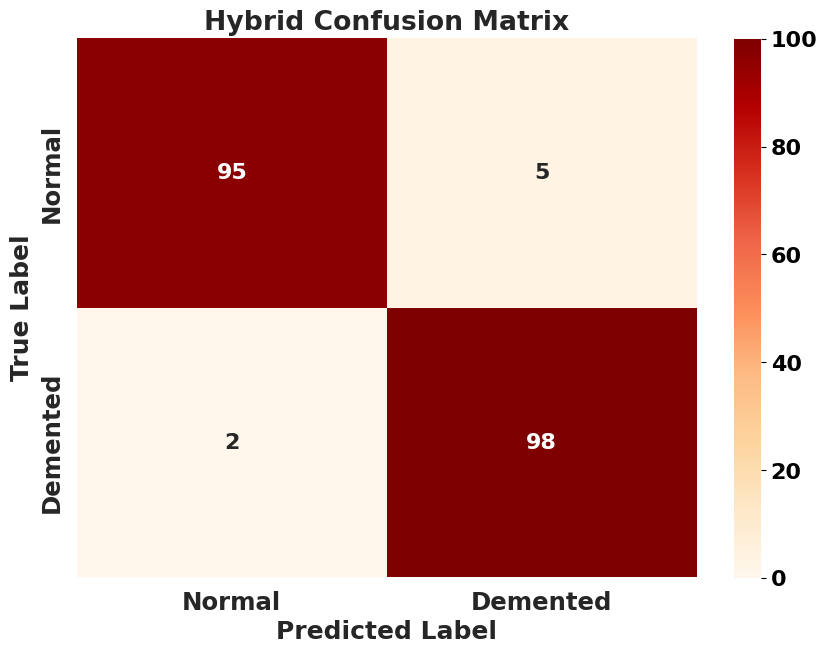}}
    \centerline{\includegraphics[scale=0.32]{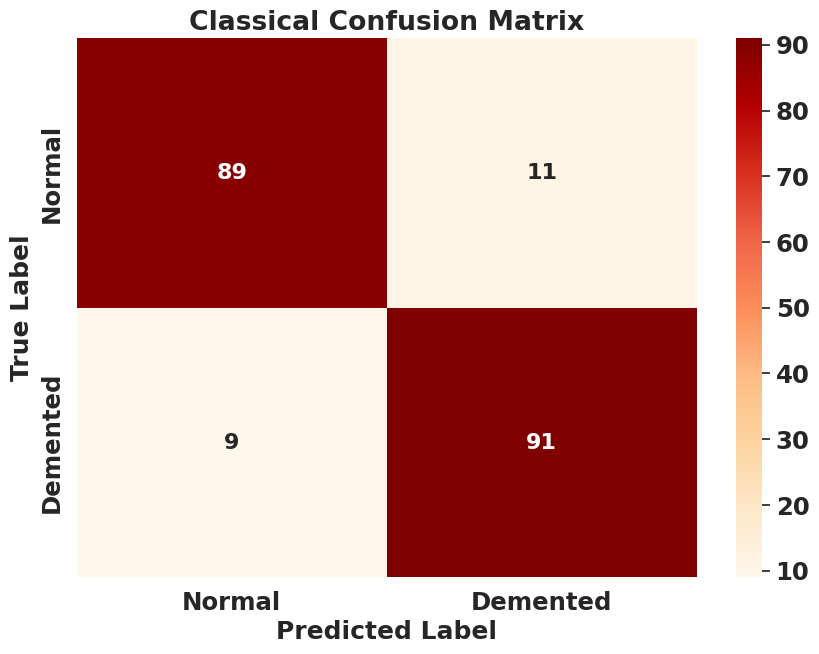}}
    \caption{Confusion matrix for demented and non-demented labels for the hybrid and classical model}
    \label{fig7}
\end{figure}

For the hybrid model testing results shown in the top matrix of Figure \ref{fig7}, out of the 100 normal images tested, 95 were correctly classified as normal and 5 were incorrectly classified as demented. Additionally, out of the 100 demented images tested, 98 were correctly classified as demented and 2 were incorrectly classified as normal. In the bottom matrix of Figure \ref{fig7}, which included the classical model testing results, 89 normal images were correctly classified as normal and 11 normal images were incorrectly classified as demented. The classical model also correctly classified 91 demented images and incorrectly classified 9 demented images as normal.

As shown in Table \ref{tab2}, the hybrid model performed better in every metric calculated: accuracy, loss, recall, precision and F1-score. Since it has a higher accuracy and F1-score, it is better at correctly predicting the right label for an image across both classes, demented and non-demented. Since the hybrid model has a better recall and precision than the classical model, it is both better at predicting correct labels out of the true positives and has a better accuracy for predicting positive labels out of the dataset.

\subsection{Statistical Analysis}
To confirm the validity of the model improvement between the hybrid model and the classic model, the McNemar's test of significance was used. The McNemar's test is often used in machine learning to compare the performance between two different binary classification models on the same dataset \cite{dietterich}. The p-value calculated indicated that the results of the study were significant.
\begin{figure*}
    \centerline{\includegraphics[scale=0.65]{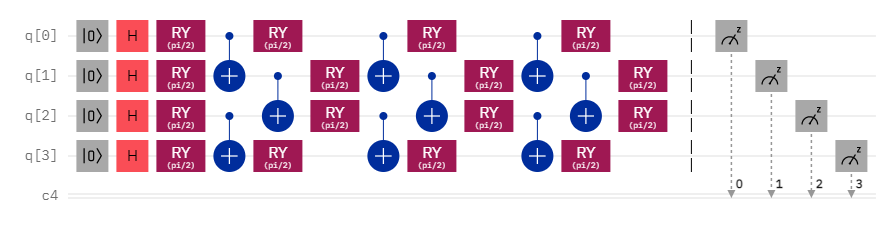}}
    \caption{IBM Quantum Composer representation of proposed variational quantum circuit architecture with a depth of 3}
    \label{composer}
\end{figure*}

\section{Conclusion}
This study found that by implementing a variational quantum circuit layer to the ResNet-18 neural network, the training accuracy, validation accuracy, training loss, validation loss, testing accuracy, recall, precision, and F1-score is improved significantly. The proposed hybrid model provides better predictions than the original ResNet-18 model due to its inherent quantum nature, which displays another example of quantum supremacy, where a quantum application provides significant speedups or benefits over a classical one \cite{harrow}. This framework of a hybrid model can be applied to other binary classification systems, as ResNet-18 is generally flexible and the variational quantum circuit can reduce a large number of parameters. A potential extension of this application could be to connect it to a video stream, and then feed in the camera frames for a real-time detection of a desired object. This hybrid model has potential to provide more accurate results for dementia from MRI diagnosis and improve patient outcomes. In the future, more layers of the proposed model can be parameterized by a quantum circuit to improve accuracy and the model can be trained on other datasets for binary classification tasks. Another improvement that could be made would be to create a separate model for each group of MRI participant (e.g. race, gender, etc.) in order to decrease the bias of the whole model.

Additionally, if ran on a remote quantum computer with enough qubits, this proposed model could potentially run faster and require less feature space than its classical counterpart. A proposed VQC of variational depth 3 and fixed angle of rotation $\pi / 2$ is shown in Figure \ref{composer}, assembled using the IBM Quantum Composer.

\section{Acknowledgment}
This research was done independently by the author, who is affiliated with the Thomas Jefferson High School for Science and Technology. No funding or mentoring was associated with this research.

\end{document}